\documentclass{iau} 
\usepackage{graphicx}

\title[Current stage of the ATCA follow-up for SPLASH] 
{Current stage of the ATCA follow-up for SPLASH}

\author[Hai-Hua Qiao, Andrew J. Walsh \& Zhi-Qiang Shen]   
{Hai-Hua Qiao$^{1,2}$ Andrew J. Walsh$^3$ \and Zhi-Qiang Shen$^{2,4}$}

\affiliation{$^1$National Time Service Center, Chinese Academy of Sciences, Xi'An, Shaanxi, China, 710600\\ email: {\tt qiaohh@shao.ac.cn} \\[\affilskip]
$^2$Shanghai Astronomical Observatory, Chinese Academy of Sciences,\\ 80 Nandan Road, Shanghai, China, 20003\\[\affilskip]
$^3$International Centre for Radio Astronomy Research, Curtin University,\\ GPO Box U1987, Perth WA 6845, Australia\\[\affilskip]
$^4${Key Laboratory of Radio Astronomy, Chinese Academy of Sciences, China}\\[\affilskip]
}
\pubyear{2017}
\volume{336}  
\jname{Astrophysical Masers: Unlocking the Mysteries of the Universe}
\editors{A. Tarchi, M.J. Reid \& P. Castangia, eds.}

\begin{document}

\maketitle

\begin{abstract}
Four ground-state OH transitions were detected in emission, absorption and maser emission in the Southern Parkes Large-Area Survey in Hydroxyl (SPLASH). We re-observed these OH masers with the Australia Telescope Compact Array to obtain positions with high accuracy ($\sim$1\,arcsec).  According to the positions, we categorised these OH masers into different classes, i.e. star formation, evolved stars, supernova remnants and unknown origin. We found one interesting OH maser source (G336.644-0.695) in the pilot region, which has been studied in detail in \cite[Qiao et al. (2016a)]{Qie2016a}. In this paper, we present the current stage of the ATCA follow-up for SPLASH and discuss the potential future researches derived from the ATCA data.
\end{abstract}

\firstsection 
\section{Introduction}

Hydroxyl (OH) was the first molecule detected at radio wavelength in the interstellar medium (ISM; \cite[Weinreb et al. 1963]{Wee1963}) and OH masers were also the first astrophysical maser species detected in the ISM (\cite[Weaver et al. 1965]{Wee1965}). Ground-state OH masers occur at 18 cm, with frequencies at 1612, 1665, 1667 and 1720\,MHz. The 1612\,MHz OH masers are usually detected toward the circumstellar envelopes of evolved stars (stellar OH masers). The 1665/1667\,MHz OH masers are mainly associated with high-mass star forming regions (interstellar OH masers). The 1720\,MHz OH masers are probes for shocks and exist in various environments, such as supernova remnants (SNRs), star forming regions (SFRs) and occasionally evolved stars. 

SPLASH (the Southern Parkes Large-Area Survey in Hydroxyl) simultaneously observed four ground-state OH transitions in an unbiased way (\cite[Dawson et al. 2014]{Dae2014}). The survey region of SPLASH is 176 square degrees including the Galactic Centre region. The spatial resolution of the Parkes radio telescope at 18\,cm is about 13\,arcmin, which is insufficient to provide positional accuracies to identify OH masers reliably. Our motivation is to get accurate positions for SPLASH OH masers using the Australia Telescope Compact Array (ATCA). The ATCA observations have been completed, which took about 330\,hours. In the SPLASH pilot region (40 square degrees), we detected 215 OH maser sites, most of which are associated with evolved stars (\cite[Qiao et al. 2016b]{Qie2016b}). The preliminary results in the Galactic Centre region (40 square degrees) showed an increase in the number of stellar OH masers and a decrease in the number of interstellar OH masers compared with the pilot region (\cite[Qiao et al. 2017]{Qie2017}). In the pilot region, we detected the 1720\,MHz OH maser emission toward a planetary nebula (G336.644-0.695; OH-maser-emitting PN; OHPN) and the magnetic fields were also measured based on the Zeeman splitting of the 1720 MHz OH masers (\cite[Qiao et al. 2016a]{Qie2016a}). The data reduction excluding the pilot region is still under way.




\section{Future plans}
{\underline{\it Polarization study}}. Based on \cite[Qiao et al. (2016a)]{Qie2016a}, ATCA data have the ability to study the Zeeman splitting of OH masers. Thus we can use these data to study the magnetic fields in both evolved star and star formation categories. Similar to the ``MAGMO'' project (\cite[Green et al. 2012]{Gre2012}), our aim is also to check whether Galactic magnetic fields can be traced with Zeeman splitting of OH masers associated with SFRs, especially for the OH masers not associated with 6.7\,GHz methanol masers. Moreover, we can also investigate the in situ magnetic fields of SNR 1720\,MHz OH masers. Zeeman splitting of evolved star OH masers will also be obtained.

{\underline{\it Maser time line in SFRs}}. We will select the star formation OH maser sites from the SPLASH survey region and compare these OH masers with other maser species, such as 6.7\,GHz methanol masers from the Methanol Multibeam survey and 22\,GHz water masers from the H$_{2}$O
southern Galactic Plane Survey. Refer to the methods in \cite[Breen et al. (2010)]{Bre2010}, we will construct a maser time line for star forming regions. 

{\underline{\it Searching for OHPNe}}. In the pilot region, 122 OH maser sites are associated with evolved stars. According to the definition of \cite[Qiao et al. (2016b)]{Qie2016b}, the 1612\,MHz spectra of 29 evolved star OH maser sites are asymmetric. Deviation from the typical double-horned profile suggests that these evolved sources may be post-AGB stars or planetary nebulae. Therefore, we can select evolved star OH maser sites with asymmetric 1612\,MHz spectra to check their continuum emission in order to determine whether these maser sites are associated with post-AGB stars or planetary nebulae.
\\
\\
\textbf{Acknowledgements:} We acknowledge the full SPLASH OH maser team who have contributed to the work presented here, i.e. Green, James A., Breen, Shari L., Dawson, J. R., Ellingsen, Simon P., G\'omez, Jos\'e F., Jordan, Christopher H., Lowe, Vicki; Jones, Paul A.. H.-H.Q. is partially supported by the Special Funding for Advanced Users, budgeted and administrated by Center for Astronomical Mega-Science, Chinese Academy of Sciences (CAMS-CAS) and CAS ``Light of West China'' Program. This work was supported in part by the Major Program of the National Natural Science Foundation of China (Grant No. 11590780, 11590784) and the Earth rotation measurement using giant fiber-optic gyroscope program.

\end{document}